# 25 Securing the Software Package Supply Chain for Critical Systems

*Ritwik Murali and Akash Ravi*

## 25.1 INTRODUCTION

The penetration of software-based systems has transformed the ways in which almost every industry operates. From controlling nuclear power stations to maneuvering spacecraft, complex software systems are used to interface with many critical systems. It is essential to ensure that these software systems are reliable and resilient. If these were to fail or get compromised, they would have a domino effect on subsequent systems. Supply chain attacks are an emerging threat targeting these systems. To quote an example of a popular widespread attack, the "SolarWinds hack" in late 2020 (Analytica, 2021) had led to a series of data breaches that affected tens of thousands of customers around the globe. Behind the screens, the cybercriminals had exploited the software package supply chain to distribute Trojan versions of the software masqueraded as updates and patches. As an example of how this attack has resulted in consequent damage, the hackers who attacked a cybersecurity firm (named FireEye) obtained unauthorized access to confidential tools that the company used for security auditing. The security flaw discovered in Apache Log4j (MITRE, 2021) is another notable vulnerability with a Common Vulnerability Scoring System (CVSS) score of 10 (the highest possible score) that had devastating consequences. The Log4j library is widely used in Java applications and thus, the vulnerability impacted a very wide range of software and services. Such vulnerabilities leave organizations exposed and susceptible to attack. More recently, Crowdstrike reported a supply chain attack on March 29, 2023, involving the popular VoIP program 3CXDesktopApp (Kucherin et al., 2023). The infection spreads through tampered 3CXDesktopApp MSI installers, including a Trojan macOS version resulting in not just financial loss, but also loss of trust for the company (Madnick, 2023).

Note that the package supply chain is not restricted only to the patches and updates. The distribution networks are involved during all stages of the software life cycle. Right from installing the tools required to set up the development environment, to pushing out newer versions of the packaged software product, different software supply chains are involved in all phases (Ohm et al., 2020). Figure 25.1 illustrates the entanglement and high involvement of software distribution supply chains when operating critical systems. This applies to various sectors like smart grids, manufacturing, healthcare, and finance. Modern infrastructure, from PLCs to data analytics, relies on multiple software systems and their supply chain dependencies. While Industry 4.0 has revolutionized processes and Industry 5.0 aims to merge cognitive computing with human intelligence, the cyber-attack surface continues to expand (Culot et al., 2019).

A software package refers to a reusable piece of software/code that can be obtained from a global registry and included in a developer's programming environment. In fact, packages serve as reusable modules integrated with developers' application code, abstracting implementation details and addressing common needs not supported by native applications, such as database connections. Most packages are available through Free and Open-Source Software (FOSS) contributions, aiding in application development by reducing time and effort. Packages may have dependencies; for example, installing package X would automatically install its dependencies like package Y. Projects may contain hundreds or thousands of dependencies managed by package managers, including those developed by the developers or published by others. For example, in the JavaScript ecosystem, the two widely employed package managers are NPM and YARN (Vu et al., 2020). CLI tools resolve packages by name and version through communication with the corresponding registry. JavaScript's popularity stems from its widespread use across the entire software and hardware stack, running on servers and mobile devices, which mutually sustains its language and package registries. As a matter of fact, in 2020, an article from the official NPM blog reported that more than 5 million developers use more than 1.3 million packages from the NPM registry, which itself caters up to 125 billion downloads every month. These statistics stand as a testimony to the popularity of package managers within developer communities.

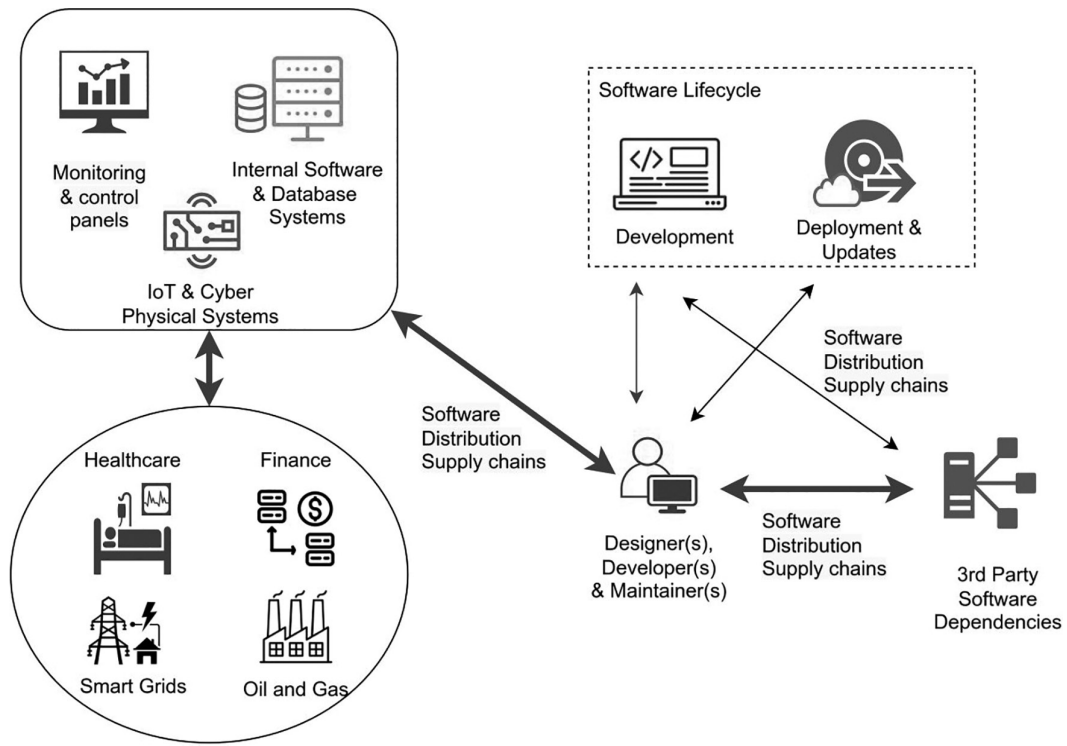

**FIGURE 25.1** Involvement of software supply chains in critical systems.

This work presents a comprehensive study of the security posture of existing package distribution (PD) systems and uses this research as a base to propose an architecture that addresses the most critical security concerns arising out of this tight coupling of software package supply chains and the infrastructure that depends on them. This proposed architecture provides end-to-end integrity of the package supply chain to mitigate the cascading effects of a critical failure. While NPM or PyPI might not be a part of the toolchain that every software developer would use, this chapter would continue to quote these systems as an indicative example of the current state of package managers. Nevertheless, the architecture itself is platform-agnostic and caters to the overall goal of securing the software package supply chains across all phases of a product's life cycle and its usage in critical systems.

Further sections discuss topics including the survey of existing studies and threat landscape analysis. Based on this, a new architecture is presented along with a demarcation of various entities and the flow of information among them. The proposed architecture can be employed to secure the acquisition of software packages, while also being used to securely distribute updates to any software. Following this, a summary of different attack vectors and corresponding mitigation strategies are also analyzed. Finally, the potential impacts of this solution are discussed before concluding the chapter.

## 25.2 RELATED WORK

### 25.2.1 STUDIES ON PACKAGE DISTRIBUTION FRAMEWORKS

With the advancement in web technologies and increased usage of web apps, there has been an exponential increase in the number of frameworks available for developers to choose from. The deployment of cloud native applications and orchestrated micro-services has also fueled the frequency and magnitude at which these services are consumed. This section hopes to present an overall survey on the current software distribution mechanisms and then analyze them in the context of critical systems to understand the threat landscape. Catering to the potential needs of web practitioners, software engineering quality metrics have been used to evaluate each alternative. Factors like modularity, scalability, and reliability play a dominant role in the perception of a framework (Graziotin and Abrahamsson, 2013). An inundation of micro-packages will result in a fragile ecosystem that becomes sensitive to any critical dependency changes. There can be a ripple effect down the dependency tree in case of any breakage (Librantz et al., 2020). Some packages perform trivial tasks, but others serve as interfaces to load foreign dependencies and third-party modules, indicating that package complexity isn't accurately defined by statistics like lines of code (LOC). Studies delve into statistics such as average package size, dependency chain size, and usage cost, emphasizing the importance of package stability and their impact on delivering end solutions (Kula et al., 2017).

The Python development ecosystem is also highly mature and growing in popularity (Bommarito and Bommarito, 2019). The repository's growth has been measured experimentally based on factors like package versions, user releases, module size, and package imports. This highlights the significance of frameworks and the extensive library availability. Enhancing PD architecture can significantly impact the IT industry, emphasizing the need for a robust and secure package manager and distribution framework. The security of these PD frameworks has been a critical concern ever since the popularity of package registries began to increase (Achuthan et al., 2014). To address the security concerns over Software Dependency Management, there have been various attempts to leverage technologies ranging from virtualization to distributed architectures (D'mello and Gonzalez-velez, 2019).

Markus Zimmermann et al. (2019) have studied the security risks for NPM users and explored several mitigation strategies. The study was performed by analyzing dependencies among packages, monitoring the maintainers responsible, and tracking publicly reported security vulnerabilities. There have also been similar attempts to devise vulnerability analysis frameworks by Ruturaj K. Vaidya et al. (2019). Once again, it is found that issues in individual packages can have a ripple effect across the ecosystem. The authors found that many projects unwittingly use vulnerable code due to lack of maintenance, even after vulnerabilities have been publicly announced for years. They compared the effectiveness of preventative techniques such as total first-party security and trusted maintainers.

When a package needs to be installed, there are a lot of tasks that happen under the hood. NPM not only downloads and extracts packages but also executes install hooks, which can include compiling sources and installing dependencies. While some tasks are essential, malicious tasks can also be run. There have been cases where post-install scripts were used to distribute malware (Wyss et al., 2022). A major incident unfolded when malicious payloads infiltrated the widely used NPM package "event-stream," impacting millions of installations. This prompted package registries to prioritize security measures. In 2018, attackers exploited systems running Electron framework apps due to outdated chromium packages, despite known vulnerabilities. NPM issued an advisory addressing a vulnerability allowing reverse shells and arbitrary data access from malicious package installations (Baldwin, 2018).

In November 2017, user "ruri12" uploaded three malicious packages – libpeshnx, libpesh, and libari – to official channels like RubyGems and PyPI, but their discovery didn't happen until July 2019 (Robert Perica, 2019). This delay prompted calls for automated malware checks. Another recent incident involved two typo-squatted Python libraries discovered stealing SSH and GPG keys (Cimpanu, 2019). Despite their removal, many developers had already incorporated them into their projects, illustrating the significant impact of such attacks on both independent developers and companies reliant on open-source frameworks and packages, causing distrust within the community. At this juncture, it is also worth pointing out that the 3CX attack (mentioned previously) was the result of another supply chain attack. A 3CX employee downloaded a tainted version of "X Trader" software in April 2022. The X Trader software was used by traders to view real-time and historical markets and developed by another company, "Trading Technologies," which discontinued the software in 2020. However, the software was still available for download from the company's website which itself was compromised in February 2022 (Page, 2023). This incident further highlights the critical nature of supply chain attacks as the potential for cascading is extremely high.

NPM offers an API to enhance visibility into the software package supply chain, providing critical information about a package's publication context. This includes metadata such as payload information, integrity hash, and Indicators of Compromise like IP addresses and file hashes. The newly introduced Security Insights API (Adam) exposes a GraphQL schema for accessing publication information. Two-factor authentication for the publishing account enhances security assessment, while publishing over the Tor network may raise suspicions of malicious behavior. Sandboxed execution and post-install script analysis can further aid in flagging tasks with malicious intent (Murali et al., 2020). For Python packages, open-source projects such as Safety DB maintain a public record of known security vulnerabilities. Packages are reviewed by filtering change logs and Common Vulnerabilities and Exposures (CVEs) for flagged keywords. However, it is worth pointing out that the vulnerabilities are only fixed after it is publicly available and not checked prior to the public announcement (Alfadel et al., 2023).

Platforms like Snyk Intel and Sonatype open-source software (OSS) index aid developers in identifying and resolving open-source vulnerabilities. The Update Framework (TUF) is a collaborative effort aimed at securing update delivery across software updaters, Library package managers, and System package managers. TUF, maintained by the Linux Foundation under the Cloud Native Computing Foundation (CNCF), safeguards compromised repository signing keys and is utilized in production systems by multiple organizations. Uptane and Upkit based on TUF guidelines have effectively secured updates for automotive and Internet of Things (IoT) devices. Despite their potential for broader application, adoption rates remain low across industries.

### 25.2.2 Current Security Landscape

To securely store and distribute packages, having accurate information is crucial for risk assessment. Current security tools often identify vulnerabilities only after an extensive audit of the end product, neglecting details about the publishing pipeline. Understanding existing mitigation methods and event flow is key to designing an effective architecture. Compromised systems offer adversaries a range of techniques to cause harm. Infected applications can exploit remote services and steal credentials. Client software vulnerabilities may expose installed packages and sensitive metadata. Adversaries can establish persistent control through malicious droppers or by connecting infected machines to a Command-and-Control (C2) server, enabling sophisticated advanced persistent threat (APT) attacks.

Attackers conduct supply chain attacks by injecting malicious code into open-source projects, targeting downstream consumers for execution during installation or runtime. They can target any project type and condition code execution based on factors like lifecycle phase, application state, operating system, or downstream component properties (Ohm et al., 2020). The attacks involve creating and promoting a distinct malicious package from scratch, entailing the development of a new open-source software (OSS) project with the intention of spreading malicious code (Balliauw, 2021). Attackers use various tactics to target users on platforms like PyPI, npm, Docker Hub, or NuGet, including promoting projects to attract victims and creating name confusion by mimicking legitimate package names. These deceptive tactics aim to trick downstream users and may involve techniques like Combosquatting, Altering Word Order, Manipulating Word Separators, Typosquatting, Built-In Package, Brandjacking, and Similarity Attack. Furthermore, attackers may subvert legitimate packages by compromising existing, trustworthy projects, injecting malicious code, taking over legitimate accounts, or tampering with version control systems to bypass project contribution workflows (Ladisa et al., 2023).

By abusing legitimate development features, malicious components can elevate privileges and move laterally through the network. Techniques such as hiding the artifacts and disabling logging mechanisms can be used to evade defenses. Most PD frameworks also have provisions to create/modify system processes. This can be utilized to execute malicious daemons and exploit system-level vulnerabilities. While one might argue that the mentioned attacks could also be performed independently, the key issue in PD frameworks (in their current form) is that they could be utilized as a trusted dropper by malicious players. Software companies are prime targets for APT actors, lacking a unified architecture to leverage knowledge from various sources for secure development. This lack can hinder traditional methods of studying adversary Tactics, Techniques, and Procedures (TTPs), enabling attack vectors to infect systems and industries using seemingly harmless software.

Looking at the current security landscape from the perspective of critical systems, the effects are even more pronounced. Recent developments in the Internet of Things (IoT) and Cyber-Physical Systems (CPS) have been revolutionizing industrial control systems (ICS) such as Supervisory Control and Data Acquisition (SCADA) networks. The integration of web and mobile applications with these systems exposes downstream systems to potential catastrophic failures due to their complex workflows and interlinked nature (Abou el Kalam, 2021). Despite hardware redundancy in most industrial deployments, software failures at key controllers could still lead to a single point of collapse. For instance, the remote manipulation of Safety Instrumented Systems (SIS) could result in severe consequences for dependent industrial facilities (Iaiani et al., 2021). State-Sponsored actors often tend to engage in warfare by compromising these systems and disrupting essential services (Izycki and Vianna, 2021). Consequently, cyber-attacks on critical infrastructure can even cost lives.

Network-based segmentation and protection are standard practices in industrial systems. However, once an adversary infiltrates a host connected to the internal network, the entire system (even if "air-gapped") becomes vulnerable. For instance, over-the-air (OTA) updates increasingly update firmware in these systems. Efforts to secure firmware updates, such as using blockchain networks, have been explored by researchers (Tsaur et al., 2022). Nevertheless, concerns persist about the security implications of computationally aided nodes (Mukherjee et al., 2021). Various protocols, including system isolation, multi-factor authentication, and integrity controls, meet security requirements. Governments mandate compliance policies, requiring training in best practices and conflict-free involvement in these systems.

Despite initiatives, inadequate scrutiny during package publication exposes a large vulnerable surface area. Users must remain vigilant regardless of project significance and seek enhanced protection against outside interference (Tomas et al., 2019). To mitigate the risks imposed by the current situation, the chapter propounds the idea of a distributed and trusted code vetting process. This work thus proposes a unified and scalable architecture that includes all stakeholders to aid users in ensuring security throughout the development process.

## 25.3 PROPOSED ARCHITECTURE

Blockchains have been regarded as a disruptive innovation that can potentially revolutionize various sectors and applications. Going by standard definitions, a blockchain is a complex data structure recording transactional records securely, transparently, and decentralized. It's a distributed ledger without a single controlling authority, open to anyone on the network. Once information is on a blockchain, it's nearly impossible to modify due to cryptographic schemes and digital signatures. Participants can reach consensus without a third party, enabling record verification. These capabilities have proven useful to establish provenance and enable key supply chain management processes (Bandara et al., 2021).

The proposed blockchain-based architecture splits the stakeholders into four different discrete entities. Publishers are those who develop packages/modules and publish them on an online repository hosted on a VCS (Version Control System) like GitHub. Package Registries index them and make these packages available to the public. Entities responsible for ensuring the security and integrity of packages are termed Observers. This would include security advisories that audit the packages and the CVE watchers who keep track of reported vulnerabilities. Finally, entities who would want to verify the security of the packages that they would be consuming are labeled as users. Depending on the context, users can be the developers who ought to download and use published packages for their projects, or users can refer to the systems deployed on a critical infrastructure that needs to verify the update packages that are being delivered to it. Figure 25.2 outlines the proposed architecture and details the interactions between the entities. In certain cases, the observers need not be external to the package registries, i.e., both these services could be provided by the same vendor. They just represent two different components.

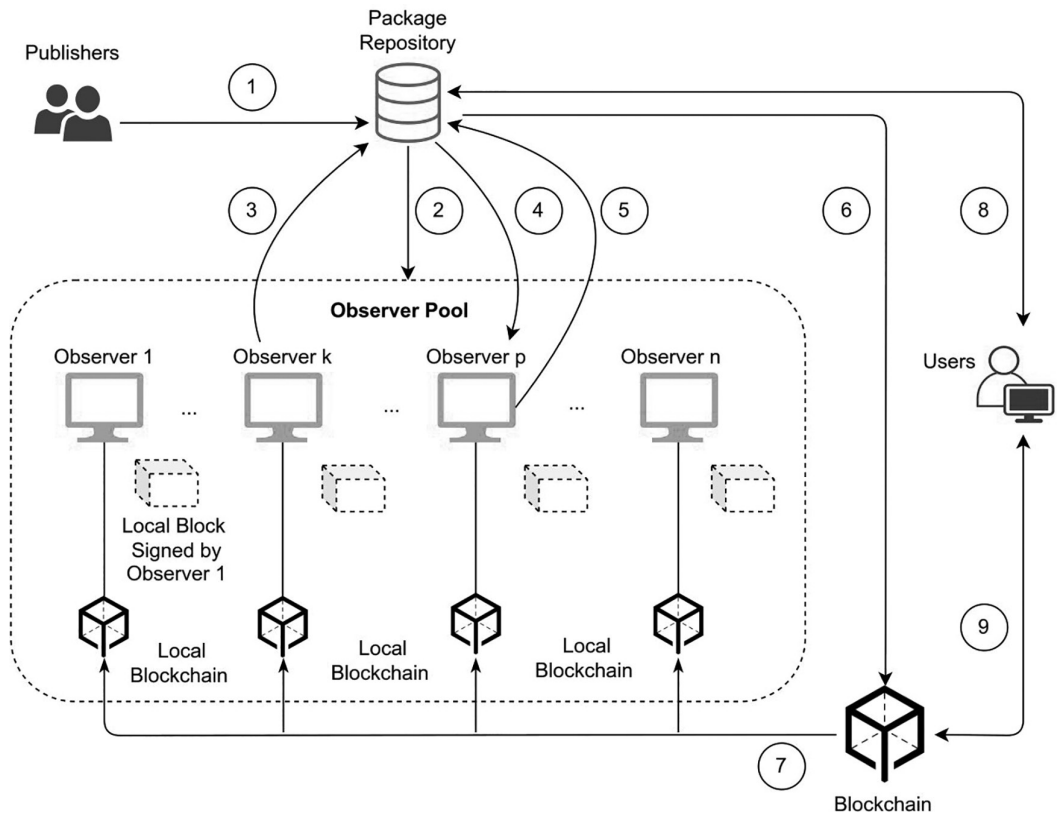

**FIGURE 25.2** Interaction between entities in proposed architecture.

Once a package has been developed and is ready for publishing, common tasks such as running tests, updating tags, and version numbers according to the Semver (Semantic Versioning) are done before pushing it to the Package Registry (Figure 25.2 Step 1). Until this step, none of the traditional methodologies needs to be modified. Once the package has been published, a copy of the package information is forwarded to all the observers in the observer pool (Figure 25.2 Step 2). They would then check if the details were authentic and if no known vulnerabilities exist. Common methods include verification of checksums and validations against VirusTotal. If the package is found to be harmless by an observer, the verification process is translated into a local block and prepared to be added to the blockchain network (Figure 25.2 Step 3). The digital asset can simply be represented as a collection of key-value pairs in binary or JSON formats. Some of the metadata that could be used to denote vulnerabilities can include a Common Vulnerability Scoring System (CVSS) score, threat classification, affected systems, etc.

Each observer accumulates their commits locally until they decide to create a block. The creation of a block would require an observer to digitally sign the proposed block using a multi-party digital signature algorithm. In addition to their private key, this scheme requires a consortium of users to sign a single blob, addressing the concerns of both group and ring signatures. Each observer would need the validation of their work from at least another co-observer which would be selected at random by the Package Registry (Figure 25.2 Step 4) who would then return the verified and signed block to the Package Registry (Figure 25.2 Step 5). Finally, the Package Registry would add this accepted block to the blockchain (Figure 25.2 Step 6). Note that adding blocks can only be performed by the Package Registry. Like how the genesis block is created in most DLTs (Distributed ledger technologies), it can be hard coded in this case also.

Since the observers can be seen as competing entities, the constant challenging of the scanning report by co-observers would result in a more accurate and accepted block. The block interval is also designed to be configurable to provide granular control over the system's functioning. Once a block is added to the chain, the observers are notified by the Package Registry to update their local copies of the blockchain with this new block based on the publicly accessible blockchain (Figure 25.2 Step 7). This process of block confirmation serves as an acknowledgment to the nodes that a proposed transaction was successfully included in the chain.

When multiple observers try to propose a block causing a race condition, the Package Registry is responsible for resolving this. The addition of blocks is done sequentially and the observer whose block wasn't added is notified to propose a new one. The resultant signature is a part of the currently accepted record and the root's final hash will be inclusive of the multi-party signature. The "previous hash" field of the next block would point to this newly computed hash and hence establish a link. This results in an immutable ledger that can securely record the verification process with federated trust management. The entire flow of data has been illustrated as a sequence diagram in Figure 25.3.

Now, when a user must download a package and include it as part of their project, the details and security of the package can be verified against the information in the blockchain network (Figure 25.2 Steps 8 and 9). The security of this architecture is enforced because every root hash is being digitally signed by multiple observers. A user who might want to verify a package would have to use the public key of the corresponding observer to read a block. Consequently, the identity of the observers is at stake which validates that the block(chain) is free of any malicious entries. This serves as a Proof of Authority (PoA) consensus algorithm that leverages the value of identities and reputations (Honnavalli et al., 2020). Algorithm 25.1 outlines the verification procedure that would be followed by the user while attempting to check if a dependency is safe to be installed.

**Algorithm 25.1: Verification of a Package Status**

> *INPUT:* Identifier for a Package that needs to be verified
> *OUTPUT:* Returns 'true' if the package is safe, else, the list of vulnerabilities is returned.
>   chainValidity():
>   *for* block in chain *do*:
>>      Check if previousHash equals to the currentHash of the previous block;
>>      *if* chain is broken *then*:
>>      *return* false;
>>      End
>   End
>   *return* true;
>   *If* chainValidity() == true *then:*
>   Find the latest block containing package information;
>   Verify the signature on the Root Hash
>   Retrieve the latest record corresponding to the concerned package and version
>   *if* package is trusted by observers *then*
>>      Initiate periodic verification of package status;
>>      *return* true;
>   End
>   End
>   *return* List of all Vulnerabilities;

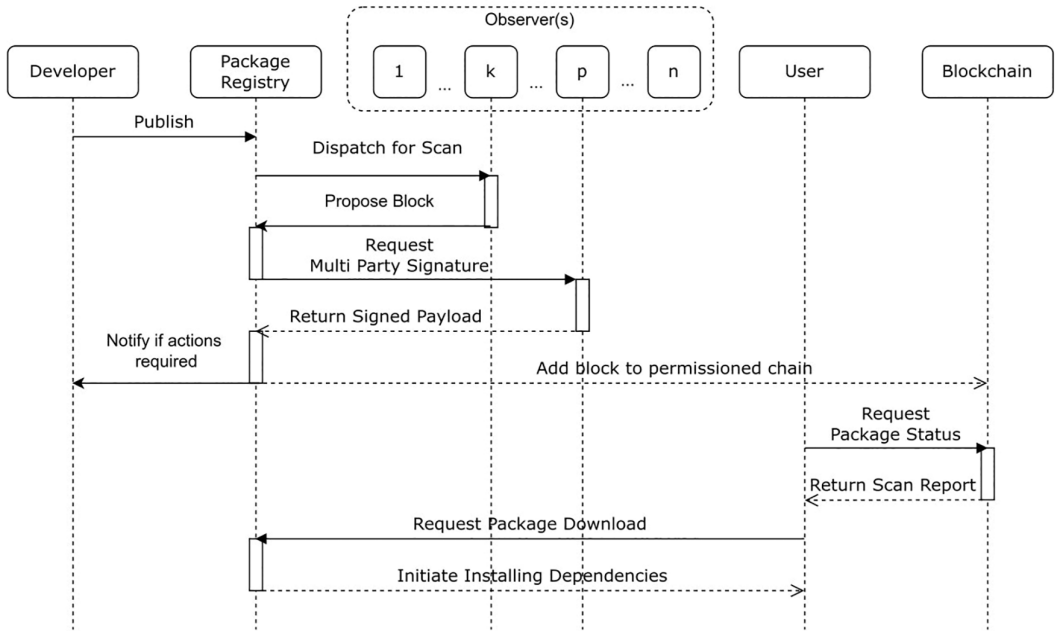

**FIGURE 25.3** Sequence diagram of the proposed architecture.

An observer would also have a numeric "rank" tagged to them. This rank would determine an observer's reputation. Each time a block is verified by a co-observer, the rank is incremented. Similarly, when the observer seems to increase false positives or true negatives while classifying threats, their rank would be downgraded. Combined with the PoA, the rank can be used to reward and penalize observers according to their participation in the network.

When multiple observers seem to have different opinions on the security of a package, an observer might decline to sign the proposed block. In this case, the Package Registry would request yet another observer to validate the block. Thus, there needs to be a minimum of two entities having the same opinion by default. However, there could be a case where only one observer was sophisticated enough to detect a threat in a package. In this case, the observer's rank can be used to determine if the block can be accepted or not. This methodology balances the occurrences of false positives and the diversity in the reporting, as each observer might report a distinct vulnerability that might have been missed by another scanner. When users are confused in choosing if a package can be trusted or not, voting ensembles can help them make informed decisions based on these insights.

The blockchain system discussed in this solution is comparable to a permissioned ledger that is open for public view and replication. Only verified observers would be allowed to add blocks to the chain via the Package Registry. All other entities would be entitled to have read-only access to this source of truth. Therefore, identity and access management controls can be effectively implemented. This brings along an array of advantages such as better scalability and faster transactions when compared to public blockchains (Ambili et al., 2017). The limited number of pre-approved block validators enables an efficient platform capable of achieving higher transactions per second (TPS). They combine the concept of "permissioning" from private associations while embracing certain principles of decentralized governance. Since there is no mining involved, recording validations are efficient and free. Such a model presents the best of both worlds and optimally addresses security concerns while balancing availability. DLTs like Hyperledger Fabric and R3 Corda can be used to construct such networks (Sajana et al., 2018). The primary reason for choosing a blockchain over any database is the requirement of needing an append-only ledger that can be read by anyone. Traditional databases do not block updates or deletes by design, which is undesirable in this use case.

When analyzing architecture in the context of securing critical systems, speed and efficiency are key aspects that would have to be ensured. Contrary to how most permissioned voting consensus systems operate, public blockchains often resort to a technique called sharding to increase transactional throughput. Fundamentally, it involves horizontally spreading out storage and computational workloads to speed up processes. In such a scenario, it would suffice if a node maintained the data related to its partition, or shard alone. In the case of the architecture mentioned in this section, explicit engineering efforts to scale up the network would not be required. Most Hyperledger implementations employing Byzantine fault-tolerant (BFT) protocols have inherent abilities to perform at scale (Sousa et al., 2018).

In a practical setting, there might be instances where the blockchain would have to "fork." They might occur due to diverging copies of the chain being maintained separately, or simply because of a software update to the system. For all the observer entities who would participate as full nodes, the same version of the processing logic must be in sync. To ensure backward compatibility with the outdated nodes, the system would ensure that soft forks are used to create an unanimously agreed consensus algorithm. In the case of most public blockchain networks, a contentious hard fork is enforced when a significant fraction of full nodes contradicts their opinion on the software versions. However, since this proposed system is designed along the lines of a permissioned ledger, this can be avoided.

Many critical systems tend to prioritize the stability of feature enhancements. Hence, software engineers writing code for such systems tend to lock the dependency versions. Being a blockchain that functions as an append-only ledger, information for the older versions is always going to be retained. Even if an update must be made for a block that has been added to the chain, it can only be added to the old one. This way, the system can also serve as an audit trail that documents all changes that have been made across versions. With various phases in which supply chains and PD networks are involved, this architecture can be used in multiple stages of the product life cycle. Being focused on interoperability, the proposed architecture builds on top of the existing stack. For effective implementation of the solution, the system doesn't require the existing framework to be replaced entirely. All the governing rules can be programmed as smart contracts based on the DLT platform of choice. This would comprise the code that contains the set of rules enforced by the system. The blockchain-based ledger can be implemented in addition to the existing system and populated asynchronously. Thus, the migration can happen gracefully and will not result in service downtime.

With full API and webhooks support, users can extend their existing workflows to work with the proposed framework. Since the entire process will be handled asynchronously, there will be no reduction in the read or write throughput of the package managers. By having periodic checks performed on the source code as a part of the CI/CD (Continuous Integration and Continuous Delivery) pipeline, organizations can verify the integrity of their development life cycle at scale. From an organizational standpoint, using this solution would lead to an agile DevSecOps cycle by introducing appropriate checks at critical stages of the software development process. Similarly, once an application is deployed, any further updates to the system can be considered a package published over an update server. In this case, the update server would be analogous to the Package Registry and all transactions can be mapped correspondingly. This way, the proposed solution can be integrated with critical systems and secure every interaction that involves pulling/pushing software.

## 25.4 ANALYSIS AND OBSERVATIONS

### 25.4.1 Security Assumptions

The proposed architecture is based on the assumption that the verdict given by the observers will be accurate to the best of their knowledge. The system assumes that the Package Registry is trusted and will not act against its functioning. Furthermore, this system does not outline the compensatory model for recognizing the commercial value of the observers. Just like the current systems where vendors have a business model where they often provide basic security and scanning services at no cost, this architecture establishes a similar environment for them to provide services. Standard security protocols need to be in place across all layers of the network stack. All communications among entities will need to happen over secure communication channels using protocols like TLS and IPsec. The certificate revocation list (CRL) will have to be checked to ensure the validity of certificate authorities (CA) and the X.509 digital certificates issued by them. This is critical to prevent man-in-the-middle attacks (MITM) and session hijacking. Attacks such as DNS (Domain Name System) cache poisoning can also be prevented by enforcing signature validation. Access control configurations need to adhere to the principle of least privilege (POLP). All server-level vulnerabilities will need to be patched and updated to prevent possibilities of security compromise and breaches. The "Blockchain Security Framework" from the OWASP Foundation could serve as a general guideline for hardening various stages of development and establishing a security baseline.

### 25.4.2 Protection Against Malicious Entities

The process of threat modeling aids in effective risk management which is critical for compliance with certain regulations and certification bodies. Here, the chapter attempts to detail the attack scenarios that have been discussed earlier and present the potential mitigation provided by the proposed architecture. The MITRE ATT&CK knowledge base has been used as a foundation for the development of threat models specific to this use case.

Scenario 1: Consider the scenario where a package has been published to a Package Registry along with an obfuscated malicious payload. These malicious commits often go unnoticed during reviewing pull requests to open-source repositories. As per the proposed architecture, once the package has been published, observers receive a trigger to evaluate the security concerns over this newly published package. Publicly known threats can be easily detected in coordination with services like VirusTotal and watching CVE listings. Regardless of whether the presence of a threat is confirmed, the scan results are recorded in the block(chain). Both the observers and these services can utilize the determined result to further enhance their datasets on which anti-malware engines are trained. The attestation of the observer is reinforced using the digital signature and the rank that is included as a part of the block's contents. Now, the observer can initiate a take-down request with the Package Registry. If any user had downloaded the malicious package, during this process, the user could securely verify the status of the package with the read-only copy of the blockchain ledger using the digital signature of the observer(s). The same verification process applies to any other package that has this malicious package as one of its dependencies. If an attacker chooses to modify the status of a package stored on the ledger, they will have to recreate the Merkle tree of the block. However, in this case, the attacker will be unable to create a valid digital signature of the block, since he does not have access to the private key of any valid observers. Assuming that a forged signature is created and put in the ledger (assuming the Package Registry is compromised), the forgery would be detected by the observers as their local blockchains would alert the stakeholders. Even if the alert is ignored, the user will still be able to detect that the data has been tampered with by verifying the identity of the entity that has signed the block. The immutability of the ledger has thus been enforced in the proposed architecture.

Scenario 2: Zero-day vulnerabilities can be discovered for packages that are already powering production systems. Initially, the threat could have gone unnoticed while the observers scanned it. The requirement is to have systems aware that they have been using a compromised package. Two features in the proposed system accommodate this requirement. First, since the ledger can have multiple blocks added to the chain corresponding to a specific package and its version, the user would have to read the latest metadata to have up-to-date information on a package. Secondly, the automated periodic verification routine on the user's end would be able to let the system know if any of the install packages have been comprised. If yes, the concerned stakeholders can be alerted to do the needful.

Scenario 3: In adverse attacks, an observer itself could be compromised and act maliciously despite their identity being held at stake. This could result in the final verdict being inverted and intentionally increase the number of false positives and true negatives. In such a case, the multi-party signature enforced by the architecture ensures that a single malicious observer cannot affect the system. Since each observer would need at least another randomly chosen entity (co-observer) to acknowledge its scan results or have a high rank based on past reputation, it becomes hard for a malicious entity to masquerade as an observer. For the entire system to be compromised, multiple observer entities will have to be controlled to successfully execute the attack. Before such a situation occurs, this behavior can be easily traced by the participating entities and their access to the permissioned blockchain can be revoked. To further harden the system, the minimum number of required co-observers can be increased at the discretion of the stakeholders. Nevertheless, this can serve as a self-regulating framework whose functioning is dictated by its stakeholders.

### 25.4.3 Advantages of the Proposed Architecture

Compared to most PD frameworks available today, the proposed architecture combines the advantages of these frameworks, while ensuring that the security concerns are effectively addressed. Essential features such as vulnerability reporting and integrity verification have been hardened by utilizing a blockchain system. The key difference is in the philosophy of enforcing security and trust. While most systems like the NPM and PyPI offer a wide distribution of trust, the proposed architecture uses a narrow distribution of trust and encourages multi-party consensus between entities that might be mutually suspicious. Based on the business requirements of organizations, the proposed solution would be able to accommodate customizable security policies and access controls on top of the core architecture. Furthermore, when inspecting this architecture with regard to critical systems, the proposed solution can be loosely integrated with legacy systems and provides graceful degradation of services in case of failures on blockchain nodes. The "zero-trust" approach ensures that every software artifact used can be verified independently. The distributed system also means that the users can offload the computational processing required at endpoints. On the technology side, the proposed system is fully compatible with proprietary protocols and data formats, eliminating concerns about vendor lock-in.

Finally, for implementing and enforcing a security measure involving multiple entities, there is an inherent need to have some commonly shared responsibilities. To incentivize the adoption of this architecture, participating entities can leverage the advantages of sharing threat intelligence (Samtani et al., 2020). All interactions happening on this system can be logged in a Security Information and Event Management (SIEM) and Security Orchestration Automated Response (SOAR) solutions for proactive monitoring and alerting. In certain cases, the collective information and statistical analysis derived from these sources can help organizations in patch management and prioritization strategies. This repository of information about the security of software packages can also serve as a source to aid Open-Source Intelligence (OSINT) and Operations Security (OPSEC).

## 25.5 CONCLUSION

Open-source developers across the globe use PD networks to publish packages and consume those shared by other contributors. Many industries and essential services are also a part of this software distribution supply chain, either as producers or as consumers. With such huge market penetration, it is not surprising that cybercriminals have started to increasingly target these systems. With the convergence of information technologies and operational technologies, attacks on the supply chain, and consequently the critical systems, can extend beyond the organization and be devastating to communities, economies, and even countries.

The blockchain-based strategy proposed in this chapter ensures the effective implementation of essential security services such as authentication, authorization, data integrity, non-repudiation, and immutability. This solution is carefully designed to be platform-agnostic and suitable for usage with various PD methodologies. Specific entities such as the package manager, users, and observers have been defined as sentinels to limit the attack surface area. The attack scenarios have been modeled considering that an attack could originate from any internal/external entity participating in the software product ecosystem. The narrow distribution of trust and multi-party consensus strategy employed by the proposed architecture ensure that the attacks are successfully mitigated. While multiple entities and transactions are mandated by the proposed architecture, the resultant system ensures that a user/developer is not delivered a piece of un-intended software that could compromise the security of the product/environment.

Due to the increasing digitization of essential infrastructure, the need for a higher level of security is quite evident. Additionally, the complexities of SCADA networks, distributed control systems, and process automation are exacerbated by the network of software dependencies their systems are relying on. The solution proposed promotes best practices and builds confidence in the PD framework by reducing the cascading impact of any failures/attacks while enhancing the security of the software package delivery supply chain.